\documentclass[prl,aps,twocolumn,showpacs,superscriptaddress]{revtex4-1}

\usepackage{lineno}
\usepackage{amssymb,amsmath}
\usepackage{natbib}
\usepackage{graphicx}
\usepackage{braket}
\usepackage{multirow}
\usepackage{color}
\usepackage[english]{babel}
\usepackage{hyperref}
\usepackage{xcolor}
\usepackage[normalem]{ulem}

\begin{document}

\title{\textbf{Dephasing and error dynamics affecting a singlet-triplet qubit during coherent spin shuttling}}

\author{Natalie D. Foster}
\email[Corresponding author: ]{ndfoste@sandia.gov}
\affiliation{Sandia National Laboratories, Albuquerque, NM 87185, USA}
\author{Jacob D. Henshaw}
\affiliation{Sandia National Laboratories, Albuquerque, NM 87185, USA}
\author{Martin Rudolph}
\affiliation{Sandia National Laboratories, Albuquerque, NM 87185, USA}
\author{Dwight R. Luhman}
\affiliation{Sandia National Laboratories, Albuquerque, NM 87185, USA}
\author{Ryan M. Jock}
\affiliation{Sandia National Laboratories, Albuquerque, NM 87185, USA}

\begin{abstract}
Quantum information transport over micron to millimeter scale distances is critical for the operation of practical quantum processors based on spin qubits. One method of achieving a long-range interaction is by coherent electron spin shuttling through an array of silicon quantum dots. In order to execute many shuttling operations with high fidelity, it is essential to understand the dynamics of qubit dephasing and relaxation during the shuttling process in order to mitigate them. However, errors arising after many repeated shuttles are not yet well documented. Here, we probe decay dynamics contributing to dephasing and relaxation of a singlet-triplet qubit during coherent spin shuttling over many $N$ repeated shuttle operations. We find that losses are dominated by magnetic dephasing for small $N<10^3$ and by incoherent shuttle errors for large $N>10^3$. Additionally, we estimate shuttle error rates below $10^{-4}$ out to at least $N=10^3$, representing an encouraging figure for future implementations of spin shuttling to entangle distant qubits.
\end{abstract}
\maketitle

\section{Introduction}
\label{sec:Introduction}

Spin qubits are a compelling candidate for the foundation of a quantum register network, having demonstrated single and two-qubit gate fidelities over 99\% \cite{Yoneda2018,Mills2022,Noiri2022,Xue2022}. By reducing the spinful $^{29}$Si nuclear background in silicon, isotopic enrichment enables coherent spin qubits to be hosted in foundry-compatible gate-defined quantum dots (QDs) \cite{Ha2022,Neyens2024,Huang2024}, streamlining integration with existing and rapidly advancing industrial silicon technology. Since spin qubit operations are carried out using short-range interactions often involving nearest neighbors \cite{Veldhorst2015,Sigillito2019}, a scalable quantum computing architecture relying on long-range quantum information transfer using spin qubits could present a challenge. One solution is to build a remote qubit registry that facilitates information exchange over processor-length scales by carrying out qubit operations at dedicated local nodes linked by interconnected hardware \cite{VanMeter2013,Vandersypen2017}. 

Linking quantum information between qubit nodes can be achieved in multiple ways. One example is by interfacing distant stationary qubit sites through an intermediate messenger, such as through virtual microwave photons to mediate remote spin-spin interactions in quantum transducers \cite{Mi2018,Borjans2020,Harvey-Collard2022}. Another approach is to physically move the spin qubit, for example, by using surface acoustic waves to form a moving QD \cite{Takada2019,Jadot2021}, or by shuttling the spin controllably through an array of gate-defined electrostatic potential traps \cite{Fujita2017,vanRiggelen-Doelman2024,Langrock2023,DeSmet2024}. In the latter, the ability to perform many repeated shuttling operations with a low error rate is key to maintaining long-range qubit entanglement. A firm understanding of error sources during shuttling is therefore required. 
For a spin qubit device composed of millions of qubits, for example, arranged on a crossbar array \cite{Li2018,Borsoi2024} or a sparse qubit node array \cite{Vandersypen2017,Kunne2024}, thousands to tens of thousands of shuttling operations may need to be performed to link sites on either end of a macroscopic processor. At present, error dynamics are not well quantified out to the large number of shuttles needed to establish these long-range interactions. 

In this work, we categorize the main error sources and dynamics during electron spin shuttling by observing a singlet-triplet qubit over many repeated coherent shuttles of a single electron spin in a gate-defined Si/SiGe triple QD device. We focus first on expected shuttling-independent loss mechanisms -- inhomogeneous dephasing and spin relaxation -- and use this baseline to classify the nature of errors dependent on the number of shuttles ($N$). From Ramsey decay measurements carried out on the stationary qubit and the qubit during shuttling, we attribute the dominant error source occurring during shuttling for $N<10^3$ to inhomogeneous dephasing arising from residual nuclear spins. We corroborate this finding by using dynamical decoupling to reduce magnetic noise and show it is independent of $N$, while also discovering the presence of nontrivial noise dynamics. From relaxation measurements, we define a bounding timescale expected of incoherent error processes in the qubit. We find, after implementing many repeated shuttling operations, that additional shuttling errors are incoherent but do not contribute appreciably to losses until $N>10^3$, meaning that errors due to shuttling are small.


\begin{figure*}[t]
\centering
	\includegraphics[width=0.75\textwidth]{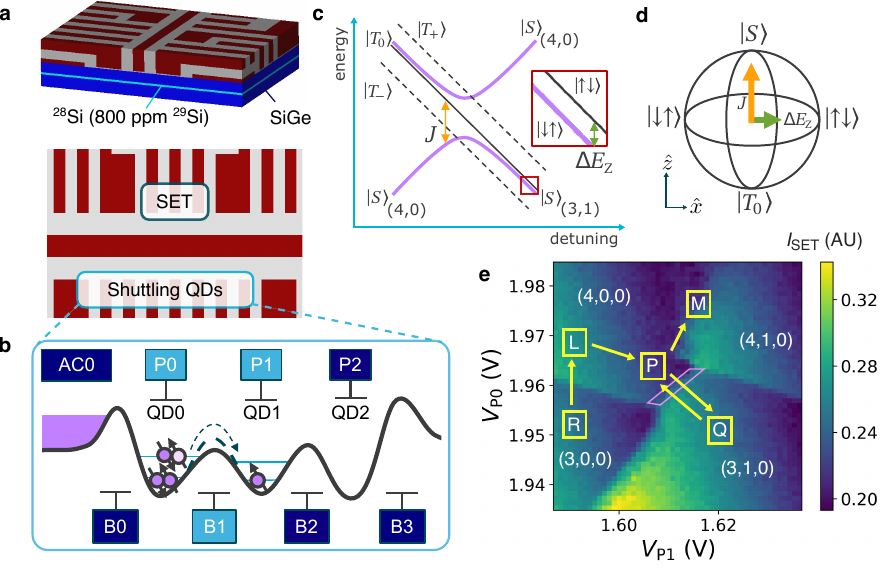}
        \caption{Singlet-triplet qubit in a triple quantum dot device. \textbf{a.} 3D profile of the device showing the $^{28}$Si quantum well (light blue) buried within SiGe (dark blue), overlapping metal gates (maroon) and oxide (gray) (above). 2D overview of the device with relevant gates composing the SET and the shuttling QDs encircled (below). \textbf{b.} Cartoon schematic of the electrostatic potential in the shuttling QDs depicting the charge transition between (4,0,0) and (3,1,0). The electron reservoir is shown under accumulation gate AC0 in purple. Electron spins ordered on energy levels (light blue horizontal lines) are moved between adjacent QDs by applying voltages to the plunger (P) and barrier (B) gates in light blue. Voltages on other gates in dark blue are nominally static. \textbf{c.} Energy diagram of the $S$-$T_0$ qubit in an applied magnetic field. $J$ is defined between $\ket{S}$ and $\ket{T_0}$ states. $\Delta E_\mathrm{Z}$ separates the $\ket{\uparrow\downarrow}$ and $\ket{\downarrow\uparrow}$ states. $\ket{\uparrow\downarrow}$ is arbitrary in permutation and cannot be distinguished from $\ket{\downarrow\uparrow}$, but is used throughout for simplicity. \textbf{d.} $S$-$T_0$ Bloch sphere corresponding to the energy diagram in \textbf{c.} showing orthogonal energy components $J$ and $\Delta E_\mathrm{Z}$ \textbf{e.} Charge stability map formed by varying P0 and P1 gate voltages showing charge transitions between QD0 and QD1. The qubit is reset in (3,0,0) for 140 $\mu$s (R), then loaded into (4,0,0) (L), rolled near the (4,0,0)–(3,1,0) PSB region (approximate location outlined in pink) to prepare for transfer (P), and then transferred by rapid adiabatic passage to (3,1,0) (Q). The spin state is measured by latched PSB at the (4,0,0)-(4,1,0) boundary (M).}
\label{fig:Fig1}	
\end{figure*}

Previous works exploring coherent shuttling span a variety of different systems ranging from enriched Si MOS \cite{Yoneda2021}, natural Si/SiGe \cite{Struck2024}, enriched Si/SiGe \cite{DeSmet2024}, Ge/SiGe \cite{vanRiggelen-Doelman2024} and AlGaAs/GaAs \cite{Fujita2017,Flentje2017} heterostructures; most were performed in external magnetic fields of about $0.2\sim1$ T and a few included micromagnets. Here, we investigate shuttling errors in the low applied external magnetic field regime ($\sim$mT) without a micromagnet, where we expect spin-orbit-induced field gradients between QD locations to be reduced. Additionally, we operate with relatively large tunnel coupling between QDs to maintain a low probability of nonadiabatic charge transfer during shuttles. Understanding the various noise mechanisms encountered during shuttling in this regime can aid the development of targeted mitigation strategies that will enable the completion of a large number of shuttles, allowing an opportunity for linking localized \cite{Pica2016,Witzel2022} qubit nodes on a sparse qubit array.

\subsection{Device overview and operation}
Fig.~\ref{fig:Fig1}a shows a schematic of the gate-defined triple QD device used in this work. Details of the device structure can be found in the Methods section. The electrochemical potential and electron occupation $n_{\mathrm{QD}i}$ at each QD$i$ are primarily controlled by voltages on plunger gates P$i$, where $i=0,1,2$ is the QD index. We denote the total charge occupation across the triple QD array by $(n_\mathrm{QD0}, n_\mathrm{QD1}, n_\mathrm{QD2})$. The tunneling rate from the electron reservoir is controlled by a voltage applied to the barrier gate B0 and the tunnel coupling between adjacent QDs is controlled by barrier gates B1 and B2 (Fig.~\ref{fig:Fig1}b).

We operate the device with four electrons in the QD chain, where two low-energy electrons under P0 form a filled-shell singlet and do not contribute significantly to the spin dynamics of the system. Fig.~\ref{fig:Fig1}c shows an energy diagram of the remaining two-spin system illustrating the relationship between the singlet ($\ket{S}$) and triplet ($\ket{T_0},~\ket{T_+},~\ket{T_-}$) states in a small magnetic field. We employ a singlet-triplet ($S$-$T_0$) qubit \cite{Petta2005}, where the qubit is encoded in the $m=0$ subspace of the 4-level system defined by linear combinations of the $\ket{\uparrow\downarrow}$ and $\ket{\downarrow\uparrow}$ states. Its logical states are defined as the eigenstates of the system in the limit of large exchange: 
\begin{equation}
\begin{aligned}
\ket{0}= \ket{S}=\frac{1}{\sqrt{2}}(\ket{\uparrow\downarrow}-\ket{\downarrow\uparrow}) \\
\ket{1}= \ket{T_0}=\frac{1}{\sqrt{2}}(\ket{\uparrow\downarrow}+\ket{\downarrow\uparrow})
\end{aligned}
\end{equation}

A schematic of the pulse sequence used to initialize and measure the qubit is depicted in Fig.~\ref{fig:Fig1}e, where we use an enhanced latching mechanism of spin-to-charge conversion through Pauli spin blockade (PSB) to measure the qubit spin state \cite{Harvey-Collard2018}. 
Two-axis control of the qubit is achieved through baseband voltage pulses to modify barrier and QD chemical potentials and toggle between control regimes dominated by the exchange interaction, $J$, and the difference in electron Zeeman energy splitting between electrons in two QDs, $\Delta E_\mathrm{Z}$ (Fig.~\ref{fig:Fig1}d). 
We use voltage pulses applied to the B1 barrier gate and adjacent P0 and P1 QD gates at the exchange symmetric operating point \cite{Martins2016,Reed2016} to achieve rotations between $\ket{\uparrow\downarrow}$ and $\ket{\downarrow\uparrow}$. 
When the electrons are well isolated and exchange is turned off, $\ket{S}$ to $\ket{T_0}$ rotations are driven by \cite{Jock2018} 
\begin{equation}
\label{eq:Zeeman_energy}
    \Delta E_\mathrm{Z} = \Delta g \mu_\mathrm{B}B,
\end{equation} 
where $g$ is the $g$-factor, $\mu_\mathrm{B}$ is the Bohr magneton, and $B$ is the magnetic field. Differences in local magnetic fields or local $g$-factors both contribute to $\Delta E_\mathrm{Z}$. The unique $B$ at each QD arises from external magnetic field gradients, or from the presence of residual spinful nuclei forming a random but slowly-varying Overhauser field \cite{Koppens2005}. The difference in $g$-factor at each QD comes from variations in interface disorder across the Si/SiGe interface giving rise to local spin-orbit interactions \cite{Jock2018}.

\section{Results}
\label{sec:Results}
\subsection{Coherent spin transfer}
\begin{figure}[t]
\centering
	\includegraphics[width=\linewidth]{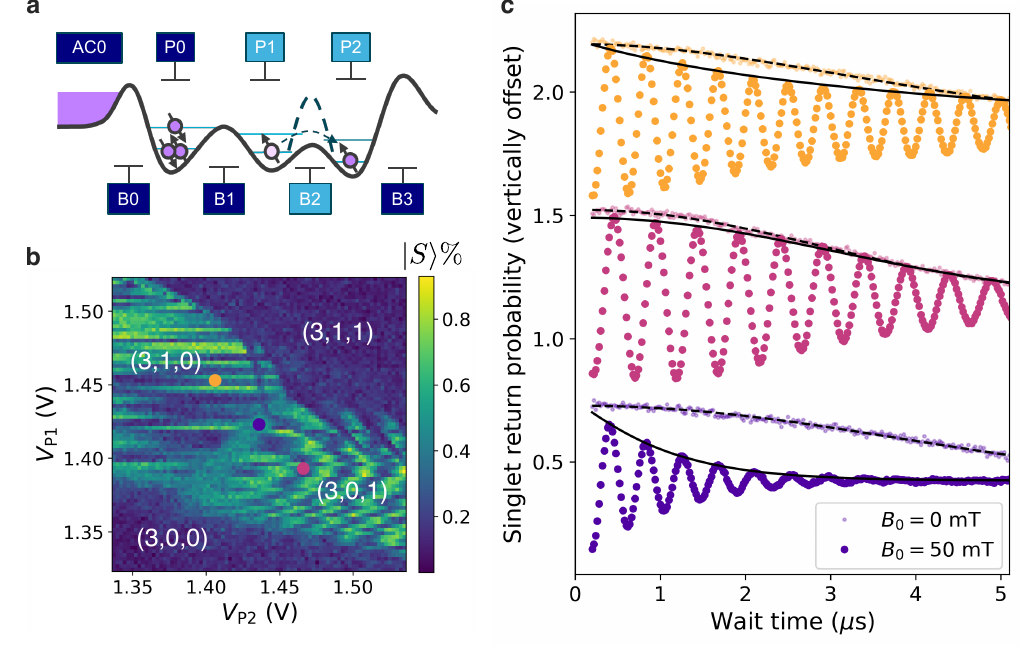}
	\caption{Shuttling and inhomogeneous dephasing. \textbf{a.} Cartoon representation of an electron being shuttled from QD1 to QD2. \textbf{b.} Singlet return probability ($\ket{S}$\%) during shuttling the detuning space surrounding the (3,1,0)-(3,0,1) charge transition. Three colored dots represent P1-P2 detunings corresponding to measurements shown in \textbf{c.} \textbf{c.} Inhomogeneous dephasing of the singlet return probability over time spent at three detunings: (3,1,0) (gold), (3,0,1) (pink) and at the charge transition (violet). Data at $B_0\approx0$ (50) mT are shown as small translucent (large opaque) circles, with decay envelope fits to Eq.~\ref{eq:exp_decay_T2s} overlaid as dashed (solid) lines. Curves are vertically offset for clarity. Full fit parameter values and uncertainties are tabulated in Supplementary Table~\ref{tab:Ramsey_fit_params}.}
\label{fig:Fig2}
\end{figure}

Two established approaches for spin shuttling are conveyor mode \cite{Seidler2022,Xue2024,Struck2024} and bucket brigade mode \cite{Flentje2017,Fujita2017,Mills2019,Noiri2022,vanRiggelen-Doelman2024,Zwerver2023}. Shuttling in this work is accomplished by a barrier pulse-assisted bucket brigade mode, in which we leverage individual gate voltage control to tailor the electrostatic potential landscape and move the spin between QDs. We prepare a $\ket{S}$ state in QD0 and QD1 and transfer one electron of the entangled spin pair from QD1 to QD2 (charge states (3,1,0) and (3,0,1), respectively) by detuning the voltages $V_\mathrm{P1}$ and $V_\mathrm{P2}$ (ramping in opposite directions), while simultaneously pulsing $V_\mathrm{B2}$ to lower the potential barrier and facilitate adiabatic transfer (Fig.~\ref{fig:Fig2}a). In principle, this shuttling method is agnostic to spin qubit encoding, provided individual spin control is possible.
The tunnel coupling between QD1 and QD2 is estimated around 52 GHz with negligible probability of completing a nonadiabatic transition (see discussion on Supplementary Fig.~\ref{fig:FigS_detuning-waittime}b for more details). We hold $V_\mathrm{B1}$ fixed to maintain negligible residual $J$ with the electron in QD0 and ensure rotations about the Bloch sphere are driven by the dominant $\Delta E_\mathrm{Z}$ interaction (see Supplementary Table~S\ref{tab:shuttling_pulse_levels} for a voltage level table).

We first demonstrate spin transfer from one QD to another by examining qubit rotations in the region near the (3,1,0)-(3,0,1) charge transition. We apply an external magnetic field of $B_0=50$ mT in the plane of the Si quantum well to intentionally drive strong $S$-$T_0$ rotations in both charge regions, showing the spin is coherently transferred.  Fig.~\ref{fig:Fig2}b shows singlet return probability following a 6 $\mu$s wait time at different QD1-QD2 detunings. This plot has four regions of interest describing different charge configurations. The bands in (3,1,0) and (3,0,1) describe differences in acquired singlet phase within each charge configuration due to changes in $g$ within each QD modifying $\Delta E_\mathrm{Z}$ as QD1 and QD2 are shaped by applied voltages, and the local interface disorder sampled by the electron varies \cite{Jock2018}. The charge transition between (3,1,0) and (3,0,1) appears as a diagonal strip of constant $\sim$50\% singlet return probability indicating a fully dephased qubit. A shift in coherent singlet return on either side of the transition indicates the spin is transferred from one isolated QD to the other (see discussion on Supplementary Fig.~\ref{fig:FigS_detuning-waittime} for more details). The low singlet return probability and absence of acquired phase bands in (3,1,1) ((3,0,0)) is due to the addition (subtraction) of an electron to the system which rapidly destroys qubit coherence. Altogether, these four delineated regions map out the charge space where spin shuttling operations will be performed and show that the entangled electron spin may be coherently transferred between QD1 and QD2.

\subsection{Spin coherence and sources of dephasing}
To demonstrate spin coherence is preserved at each shuttle step and to quantify noise dynamics that drive dephasing independent of shuttling, we perform a Ramsey decay experiment by preparing a $\ket{S}$ state and letting it evolve under $\Delta E_\mathrm{Z}$. We measure the singlet return probability over time, averaged over 1 hour total to ensure the ergodic limit was reached \cite{Madzik2020} (Supplementary Fig.~\ref{fig:FigS_T2star}). Fig.~\ref{fig:Fig2}c shows the singlet return at different locations on the charge stability map in Fig.~\ref{fig:Fig2}b indicated as colored circles corresponding to (3,1,0), (3,0,1), and the charge transition. 

We first consider dephasing in the case of negligible external magnetic field ($B_0\approx0$ mT, with a remanent field of up to a few mT due to trapped flux in the superconducting magnet) such that dephasing is expected to be dominated by fluctuations in the local Overhauser field \cite{Petta2005}, where no $S$-$T_0$ rotations are observed. We fit the data to a decay profile of the form
\begin{equation}
    f_{T_2^*}(t) = Ae^{-\left(t/T_2^*\right)^\alpha}+y_0,
\label{eq:exp_decay_T2s}
\end{equation}
where $A$ is the amplitude, $y_0$ is the vertical offset, $T_2^*$ is the inhomogeneous dephasing time, and $\alpha$ is the decay exponent. The resulting dephasing times and fit uncertainties are $T_2^*(3,1,0)=4.60\pm0.02~\mu$s, $T_2^*(3,0,1)=3.86\pm0.02~\mu$s, and $T_2^*(\mathrm{transition})=5.12\pm0.03~\mu$s, and are consistent with previously measured quantities for QDs in 800 ppm $^{29}$Si \cite{Witzel22012, Eng2015, Neyens2024}. The variation among $T_2^*$s suggests local sources of dephasing affect the spin in each QD, including the wavefunction overlap with nearby $^{29}$Si and $^{73}$Ge. We interpret the longer $T_2^*(\mathrm{transition})$ as the result of a process like motional narrowing -- effectively averaging out magnetic noise acquired from the different nuclear distributions when the spin is hybridized between QD1 and QD2 \cite{Mortemousque2021}. This increase in $T_2^*$ when the second electron interacts with a larger number of nuclear spins when hybridized is consistent with the central limit theorem, which scales as $T_2^*\sim\sqrt{N_\mathrm{spins}}$, where $N_\mathrm{spins}$ is the number of spinful nuclei lying within the electron wavefunction \cite{Witzel2012}. The resulting decay exponent fits, $\alpha(3,1,0)=1.92\pm0.02$, $\alpha(3,0,1)=2.16\pm0.02$, and $\alpha(\mathrm{transition})=2.12\pm0.02$, resemble Gaussian profiles as expected of inhomogeneous nuclear spin dephasing \cite{Petta2005}. 

Next, we apply a small external field of $B_0=50$ mT to amplify and study the effect of charge noise acting on $\Delta E_\mathrm{Z}$. Clear $S$-$T_0$ oscillations are observed and fit to a decaying sine function of the form
\begin{equation}
    f_{T_2^*}(t) = Ae^{-\left(t/T_2^*\right)^\alpha}\sin\left(2\pi f t\right)+y_0,
\label{eq:dephase_sine}
\end{equation}
where $f$ is the $S$-$T_0$ rotation frequency. The resulting dephasing times are $T_2^*(3,1,0)=3.82\pm0.12~\mu$s, $T_2^*(3,0,1)=4.09\pm0.03~\mu$s, and $T_2^*(\mathrm{transition})=1.08\pm0.01~\mu$s. 
We attribute the fast decay observed at the charge transition to the hybridization of the electron wavefunction split between QD1 and QD2 \cite{Yoneda2021,vanRiggelen-Doelman2024}, which enhances the qubit sensitivity to charge noise. Here, charge noise rapidly alters the extent of the wavefunction overlap within QD1 and QD2, each with its own distinct spin-orbit interactions. This manifests as increased fluctuations in $\Delta E_\mathrm{Z}$ experienced by the qubit, which increases with magnetic field, driving faster $S$-$T_0$ dephasing. This sensitivity to charge noise may be minimized by tuning the external magnetic field angle \cite{Tanttu2019}, or by simply reducing the magnitude. For the remainder of this work concerning shuttling, we maintain a small or negligible field to reduce the susceptibility of the qubit to noise through $\Delta E_\mathrm{Z}$.

Additionally, we observe changes to $\alpha$ that imply changes to the noise spectrum. The resulting decay exponent fits are $\alpha(3,1,0)=0.90\pm0.06$, $\alpha(3,0,1)=2.12\pm0.06$, and $\alpha(\mathrm{transition})=1.14\pm0.02$, ranging from exponential to Gaussian forms. Since the noise spectrum and mechanisms dictate the observed decay profile \cite{Stano2022}, the various $\alpha$ suggest different localized noise dynamics, in addition to residual nuclear spins, contribute to dephasing in each QD.
One possible source known to exhibit non-Gaussian decay behavior is random telegraph noise originating from charge traps, where the observed charge noise is often attributed to two-level systems (TLSs) that are distributed nonuniformly within the gate dielectric \cite{Connors2019,Struck2020}. 
This is consistent with our assessment that charge noise dominates the fast $T_2^*(\mathrm{transition})$, and suggests an increased susceptibility to charge noise in QD1 compared to QD2, since $\alpha(3,1,0)$ is exponential and $\alpha(3,0,1)$ is Gaussian. Overall, we observe qubit inhomogeneous dephasing to be primarily driven by fluctuations in the Overhauser fields generated by residual nuclear spins near 0 mT, and a combination of nuclear spins and charge noise at 50 mT.

\begin{figure*}
\centering
	\includegraphics[width=0.75\textwidth]{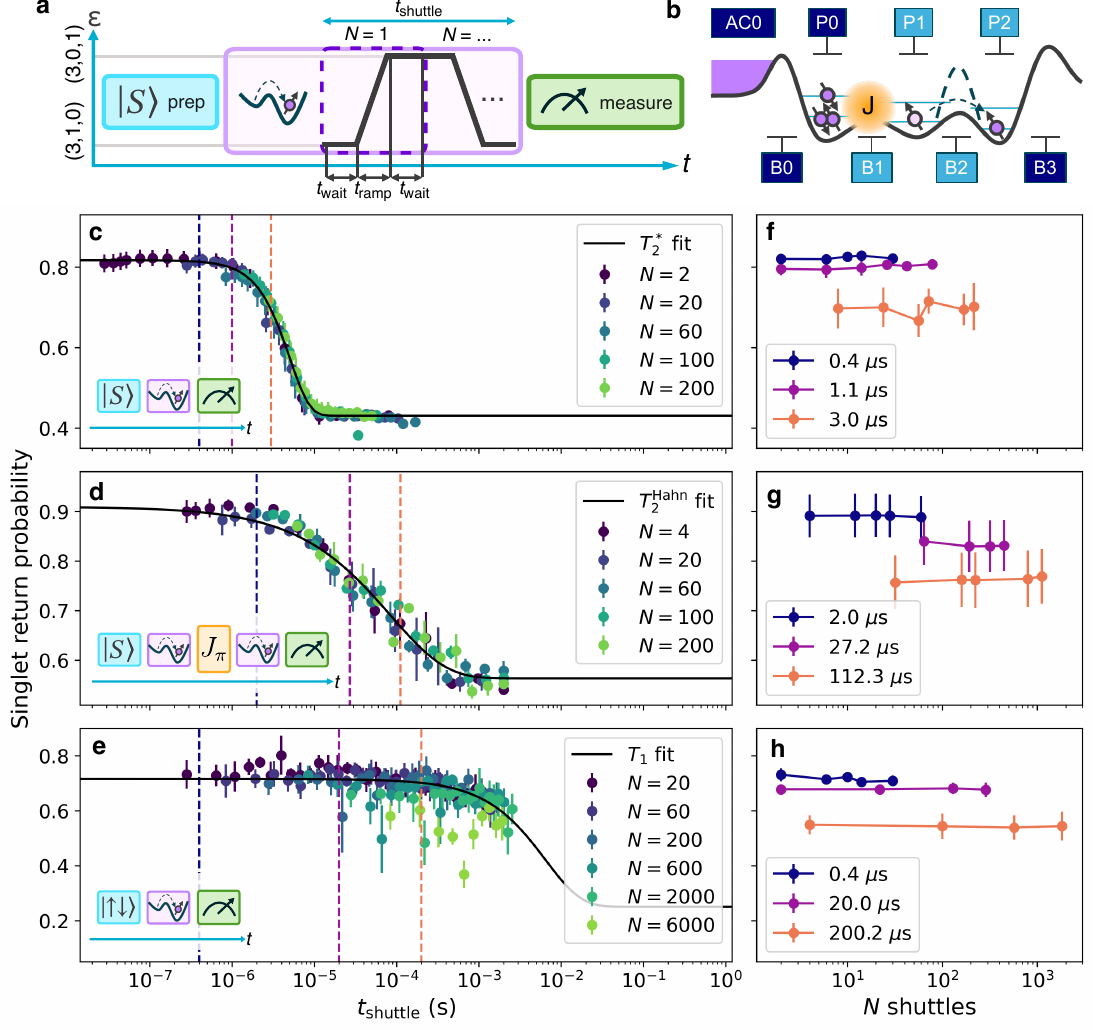}
	\caption{Repeated shuttling experiments. \textbf{a.} Pulse sequence diagram for repeated shuttles. A $\ket{S}$ state is prepared, undergoes shuttling between QD1 and QD2, and then its final spin state is measured. \textbf{b.} Cartoon representation of the exchange interaction between QD0 and QD1,2 used to carry out a $\pi_z$ pulse for dynamical decoupling. \textbf{c.} Singlet return probability after $N$ shuttles (different colored markers) are executed over a range of time. A fit of the form of Eq.~\ref{eq:exp_decay_T2s} is overlaid (black curve). \textbf{d.} Singlet return probability for an experiment similar to \textbf{c.}, but with the dynamical decoupling sequence shown in the inset applied. A fit of the form of Eq.~\ref{eq:Hahn_decay} is overlaid (black curve). \textbf{e.} Singlet return probability for a $\ket{\uparrow\downarrow}$ state after $N$ shuttles executed over time. An exponential decay fit of the form of Eq.~\ref{eq:exp_decay} is overlaid (black curve). \textbf{f., g., h.} Repeated shuttles carried out at fixed $t_\mathrm{shuttle}$ corresponding to the colored vertical dashed lines in \textbf{c., d.,} and \textbf{e.}, respectively. $B_0\approx0$ mT for the $\ket{S}$ shuttles (c.,d.,f.,g.) and $B_0=10$ mT for the $\ket{\uparrow\downarrow}$ shuttles (e., h.). All error bars shown represent the standard deviation of the mean of the measurement.
 }
\label{fig:Fig3}
\end{figure*}

\subsection{Repeated spin shuttles}
To understand the role of shuttle-induced errors, independent of intrinsic noise, we perform repeated spin shuttling operations to investigate errors from shuttling and observe the singlet return probability. We implement repetitions of the procedure outlined in the previous section to transfer an electron back and forth between QD1 and QD2. The voltage pulses are structured such that the shuttled electron spends an equal amount of time in each QD during shuttling ($2t_\mathrm{wait}$), and the ramp time ($t_\mathrm{ramp}$) is fixed at 10 ns between charge configurations to ensure adiabatic charge transfer. A single shuttle is defined as one transfer of an electron between QD1 and QD2 (from either of (3,1,0) or (3,0,1) to the other), where the total time spent shuttling $N$ times is $t_\mathrm{shuttle}=N(2t_\mathrm{wait}+t_\mathrm{ramp})$ (Fig.~\ref{fig:Fig3}a).

To investigate dephasing during shuttling, we vary $t_\mathrm{wait}$ for a series of fixed $N$ carried out within $t_\mathrm{shuttle}$ (Fig.~\ref{fig:Fig3}c), with $B_0\approx0$ mT to minimize the charge noise-induced field-dependent dynamics on $\Delta E_\mathrm{Z}$ discussed above. We observe a single decay profile in singlet return, independent of $N$, indicating the decay is primarily time-dependent. A decay fit of the form in Eq.~\ref{eq:exp_decay_T2s} to the combined data reveals $T_2^*=5.07\pm0.07~\mu$s, which is consistent with $T_2^*(\mathrm{transition})$ measured in the absence of shuttling at $B_0$ $~=0$ mT (Fig.~\ref{fig:Fig2}c). This indicates the spin samples a similar noise environment when shuttled and hybridized. This observation suggests the error introduced by the act of shuttling is low. The decay profile is nearly Gaussian with $\alpha=1.82\pm0.07$, which points to the dominant dephasing mechanism being nuclear spin fluctuations.

Next, we employ a standard spin echo technique \cite{Petta2005} to reduce the effect of slow magnetic noise and extend the qubit dephasing time. This allows more shuttle operations to be performed within the qubit coherence time, to further highlight the role of shuttle-induced errors. For this experiment, $B_0\approx0$ mT to minimize field-dependent effects on $\Delta E_\mathrm{Z}$. We prepare the $\ket{S}$ state and allow the two electron spins to evolve while one is stationary in QD0 and the other is repeatedly shuttled between QD1 and QD2 $N/2$ times. A calibrated $J$ pulse on B1 applies a $\pi_z$ operation to swap the two spins (Fig.~\ref{fig:Fig3}b), followed by $N/2$ shuttles such that both spins sample the same magnetic environment for an equal amount of time, negating any acquired phase difference due to nuclear spin fluctuations slower than $1/2t_\mathrm{shuttle}$. Fig.~\ref{fig:Fig3}d shows the singlet return probability after implementing the spin echo sequence during shuttling for a range of $N$ completed within $t_\mathrm{shuttle}$. We observe a single decay profile independent of $N$ and fit the data to a generic decay function of the form
\begin{equation}
    f_{T_2^\mathrm{Hahn}}(t) = Ae^{-\left(t/T_2^\mathrm{Hahn}\right)^\alpha}+y_0,
\label{eq:Hahn_decay}
\end{equation}
similar to Eq.~\ref{eq:exp_decay_T2s}. The fit yields $T_2^\mathrm{Hahn}=84.48\pm7.82~\mu$s which is reasonably consistent with previously measured values in Si/SiGe heterostructures \cite{Kawakami2014,Kerckhoff2021}, and indicates sources of noise contributing to dephasing have been reduced. The fit converges with an exponent of $\alpha=0.64\pm0.06$, which deviates from $\alpha$ commonly observed in the range of $1\lesssim\alpha\lesssim2$ \cite{Petta2005,Kawakami2014,Veldhorst2014}, suggesting a different noise spectrum is introduced. Sub-exponential $T_2$ decays have been previously observed \cite{Stanwix2010} and simulated \cite{Schatzle2024} in other semiconductor-based spin systems, such as NV$^-$ centers in diamond, with origins tracing back to the particular spin dynamics of the bath. In our system, the exact mechanisms driving decay are as of yet unknown but may be related to bath dynamics at low magnetic field.

Next, we investigate the role of spin relaxation during shuttling since spin-flip mechanisms are known error sources in Si QDs \cite{Yang2013}. We prepare an eigenstate $\ket{\uparrow\downarrow}_{(3,1,0)}$ of the $\Delta E_\mathrm{Z}$ interaction and monitor return of this state while repeatedly shuttling between QD1 and QD2. We use a small applied field of $B_0=10$ mT to avoid mixing with $\ket{T_-}$ during the (3,1,0) charge state preparation. Fig.~\ref{fig:Fig3}e shows singlet return probability of the shuttled $\ket{\uparrow\downarrow}$ state for a range of $N$ completed within $t_\mathrm{shuttle}$. We observe a single decay profile independent of $N$ and fit the data to an exponential decay function of the form
\begin{equation}
        f_{T_1}(t) = Ae^{-t/T_1}+y_0
\label{eq:exp_decay}
\end{equation}
with a range of assumptions placed on $y_0$, the singlet return saturation, since it is not explicitly measured. The fit results in $T_1=6.22\pm0.71$ ms, where the uncertainty represents fit variation based on $y_0$, and lies within the broad range of $T_1$ measured for electrons confined to Si QDs, typically falling between milliseconds and seconds \cite{Petit2018,Eng2015,Borjans2019}. A variety of mechanisms generally stemming from spin-orbit and spin-valley interactions contribute to spin relaxation in Si QDs \cite{Tahan2014}. Due to the natural disorder and inversion asymmetry at the Si/SiGe interface, the small in-plane $B_0$ gives rise to spin-orbit effects \cite{Jock2018} that can induce spin-flip relaxation over time. At $B_0=0\sim10$ mT, we expect a small but nonzero influence from the spin-valley hotspot measured near 190 mT in QD0, corresponding to a valley-splitting energy of $\sim$21 $\mu$eV in QD0 (see discussion on Supplementary Fig.~\ref{fig:FigS_Bramp-FFT}), which could drive a faster relaxation rate ($T_1^{-1}$) \cite{Yang2013}. Spin-valley hotspots for the other QDs are higher in energy and would likely play a smaller role. Few references for $T_1$ in the low-field limit exist; however, the observation that $T_1>T_2^\mathrm{Hahn}$ in our device shows that incoherent errors due to spin-flip processes are relatively small compared to coherent errors arising from magnetic noise during shuttling for $N\lesssim10^3$.

From the data presented in Figs.~\ref{fig:Fig3}c, d, and e, we anticipate the shuttling error to be low since the time-dependent singlet return decays in each panel are nearly indistinguishable among different $N$. To assess the time-independent error due to the shuttling operation, we fix $t_\mathrm{shuttle}$ while varying $N$ such that the effect of time-dependent dephasing and errors is constant with $N$. Figs.~\ref{fig:Fig3}f, g, and h show the measured singlet return probability at three fixed $t_\mathrm{shuttle}$ corresponding to the shuttling experiments shown in the insets of Figs.~\ref{fig:Fig3}c, d, and e, respectively. The error bars represent standard deviations of the mean of singlet return spanning repeated measurements totaling at least one hour. In all cases, the decay in singlet return is undetectable within error bars, and suggests shuttle errors are negligible out to at least $N\approx10^3$ (Fig.~\ref{fig:Fig3}g). We make an attempt to quantify the worst-case estimate of shuttling error by fitting a simple linear function from the top of the first error bar to the bottom of the last error bar and extracting the slope of the line (not shown), which bounds the error per shuttle below $10^{-4}$, an improvement from recently reported shuttling errors.

\begin{figure}[t]
\centering
	\includegraphics[width=\linewidth]{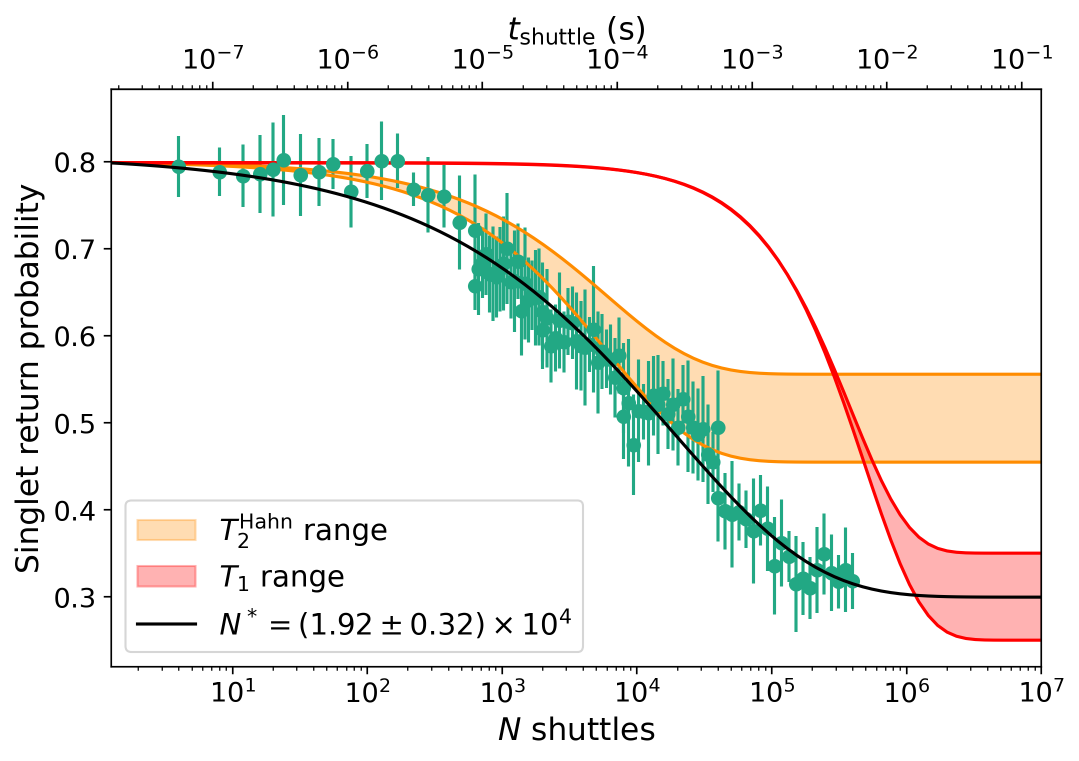}
	\caption{Shuttling error estimates at large $N$. Data is shown in teal markers where error bars represent standard deviations of the mean of multiple repeated measurements. A phenomenological fit of the form of Eq.~\ref{eq:phenomenological} is overlaid (black curve). The range of $T_2^\mathrm{Hahn}$ ($T_1$) based on analysis in the main text regarding Fig.~\ref{fig:Fig3}d. (Fig.~\ref{fig:Fig3}e.) is shown in orange (red) shading, presented as a guide to the eye. The top horizontal axis ($t_\mathrm{shuttle})$ is related to the bottom ($N$ shuttles) by the expression $t_\mathrm{shuttle}= N(2t_\mathrm{wait}+t_\mathrm{ramp})=14N~\mathrm{ns}$. Here, $B_0\approx0$ mT.}
\label{fig:Fig4}
\end{figure}

Due to the low shuttling error, many repeated $N$ are required to identify and elucidate the nature of shuttle-induced losses. To do so we extend the previously discussed spin echo shuttling technique out to the maximum $N$ possible within our instrument bandwidth capabilities. We compose a shuttling experiment at $B_0\approx0$ mT, with the minimum possible $2t_\mathrm{wait}=4$ ns and increase the repeated shuttles out to $N=4\times10^5$, spanning a total time of $t_\mathrm{shuttle}=5.6$ ms, where $t_\mathrm{shuttle}(N)=14N$ ns. The singlet return probability over $N$, and equivalent $t_\mathrm{shuttle}$, is shown in Fig.~\ref{fig:Fig4}. We consider the observed decay in the context of estimated $T_2^\mathrm{Hahn}$ and $T_1$ dynamics in this device, which are presented as shaded colored regions and are scaled by a multiplicative factor accounting for SPAM variation, serving as a guide to the eye. For a small number of shuttles $N<10^3$, the $T_2^\mathrm{Hahn}$ decay profile agrees with the data and implies the decay is dominated by the nontrivial magnetic spectral noise acting on $\Delta E_\mathrm{Z}$. For a large number of shuttles $N>10^3$, the decay deviates from this behavior and appears to be dominated by a different mechanism. The observation that the singlet return saturates below 50\% indicates that the shuttling error is due to an incoherent spin relaxation process. However, the process causes decay on a timescale faster than the measured $T_1$ (for $N<10^3$), indicating a shuttle-induced, or $N$-dependent mechanism dominating spin relaxation when shuttling a large number of times ($N>10^3$). 


To characterize the nature of the shuttling error as a function of the number of shuttles, we fit the data to a phenomenological function of the form  
\begin{equation}
    f_{N^*}(N)=Ae^{-(N/N^*)^\alpha}+y_0, 
\label{eq:phenomenological}
\end{equation}
which defines the characteristic decay in units of $N$ ($N^*$). We find $N^*=(1.92\pm 0.32)\times10^4$ shuttles can be completed within a shuttle error-limited characteristic decay period, 1-2 orders of magnitude more than recently demonstrated with shuttled spins \cite{vanRiggelen-Doelman2024,DeSmet2024}. The observation that the function fits poorly for smaller $N<10^3$ suggests the influence of shuttle errors do not become apparent until after the completion of $N\approx10^3$ shuttles. The fit results in $\alpha=0.41\pm0.03$ and, although a single decay profile is insufficient to describe the losses observed during shuttling, it points to the complexity of competing noise and error mechanisms at play during shuttling as $N$ increases. 

\begin{table*}[t!]
\centering
\begin{tabular}{l|l|l|l|l}
Shuttle experiment     & Decay fit             & Lifetime ($\tau$)           & Exponent ($\alpha$)  & Error per shuttle ($p$)  \\ \hline
Ramsey                 & Gaussian              & $T_2^*=5.07\pm0.07~\mu$s            & $1.82\pm0.07$    & $2.2\times10^{-5}$      \\
Relaxation                     & Exponential           & $T_1=6.22\pm0.71$ ms                & 1 (fixed)   & $2.3\times10^{-6}$      \\
\end{tabular}
\caption{Shuttle errors based on decay fit calculated from the time it takes to complete one shuttle ($p=1-e^{-(14~\mathrm{ns}/\tau)^\alpha}$)}
\label{tab:fit_params}
\end{table*}

\section{Discussion}

Understanding sources of error during shuttling is essential to the development of a successful spin shuttling platform used to entangle remote qubits. In this work, we have built on previous efforts exploring bucket brigade mode shuttling in Si \cite{Yoneda2021,Noiri2022} and in Ge \cite{vanRiggelen-Doelman2024} QDs. Using an enhanced barrier pulse-assisted approach, we have demonstrated coherent spin shuttling in the limit of a large number of shuttles, with the goal of categorizing the different dephasing and relaxation mechanisms affecting the qubit while shuttling. 
Independent of shuttling, we identify magnetic noise from Overhauser field fluctuations as the dominant source of dephasing in the absence of external fields ($B_0\approx0$ mT) and a combination of nuclear bath noise and charge noise at finite field ($B_0=50$ mT). When the spin is repeatedly shuttled, we observe little change in the decay at $B_0\approx0$ mT, indicating that shuttling operations marginally contribute to losses. Using a standard spin echo technique, we show the coherence time during shuttling may be extended owing to the reduction of magnetic noise on $\Delta E_\mathrm{Z}$ and is limited by a different noise spectrum related to hyperfine interactions indicated by a subexponential $T_2^\mathrm{Hahn}$ decay profile. Residual uncorrelated magnetic noise on $T_2^\mathrm{Hahn}$ timescales is determined to be the limiting time-dependent contribution since $T_2^\mathrm{Hahn}<T_1$. Additionally, we find that $T_2^*$, $T_2^\mathrm{Hahn}$, and $T_1$ appear to be $N$-independent in the small $N<10^3$ regime and do not contribute appreciably to shuttling error.
Fitting the dephasing and relaxation data in this regime, (Figs.~\ref{fig:Fig3}f., g., h.), we make a rudimentary yet encouraging shuttle error estimate of $10^{-4}$.

By executing shuttle operations in the large $N$ regime with a maximum of $N=4\times10^5$ shuttles, we identify a shuttling-induced incoherent error which contributes significantly only after many shuttle operations. In the small $N$ ($<10^3$) regime, shuttling errors are negligible and decay is primarily limited by magnetic dephasing on $T_2^\mathrm{Hahn}$ timescales. In the large $N$ ($>10^3$) regime, a shuttle-induced, or $N$-dependent mechanism driving spin relaxation causes decay, which occurs on a timescale faster than the measured $T_1$ (for $N<10^3$). 
The exact mechanisms driving shuttle errors are unclear given the evidence, but their origins can be narrowed down. Nonadiabatic charge transfer is expected to contribute imperceptibly small errors due to the large tunnel coupling between QD1 and QD2, as previously noted. We also expect errors arising from electron-nuclear spin flip-flops when repeatedly ionizing QD1 and QD2 to be low ($<10^{-4}$) \cite{Witzel2022}.

On the other hand, the physical motion of the electron through the transition may give rise to additional relaxation channels.
Charge noise is known to induce spin relaxation through driven transverse spin-orbit fluctuations, which would be most pronounced at the QD charge transition \cite{Borjans2019}. Furthermore, the strong charge hybridization at the transition could lead to an effective reduction in valley or orbital splitting as well as changes in the spin-valley coupling, which could enhance spin relaxation though spin-valley hotspot-like mechanisms \cite{Yang2013}. Reduced valley splitting or rapid change of valley phase when shuttling between QDs may also increase susceptibility to excitations outside of the valley ground state during transfer, but recent theoretical work suggests encouraging evidence that optimized control schedules may reduce this likelihood during spin shuttling \cite{Lima2024}.
In the above scenarios, the probability of these spin-flip errors occurring would increase as more shuttles are appended, or as the shuttled electron spends more time near the charge transition. While these values have not been measured explicitly in the low magnetic field regime, they can be targeted as the focus of future studies expanding upon spin shuttling errors.

Looking ahead, error estimates based on the $T_2^*$ and $T_1$ decays are the more practical expectation for building up a long-range spin shuttling platform. Focusing on the timescales quantified in this work ($T_2^*$ and $T_1$), we make estimates of the error per shuttle $p=1-e^{-(14\mathrm{ns}/\tau)^\alpha}$ based on the assumption of the limiting decay timescale $\tau$ during a 14 ns shuttling operation, summarized in Table~\ref{tab:fit_params}. If all other noise sources were remediated and the fluctuating field gradient is the only remaining noise mechanism so that decay is limited by $T_2^*$, we extrapolate the shuttling error to be $p=2.2\times10^{-5}$. If noise on $\Delta E_\mathrm{Z}$ could also be reduced to the level such that spin-flip relaxation on the $T_1$ timescale limits decay, the shuttle error is estimated even lower at $p=2.3\times10^{-6}$. These results highlight the potential for high-fidelity shuttling as a promising pathway to remote spin entanglement in spin-based quantum computing platforms.

\section{Methods}
\subsection{Device details}
The device was fabricated by Intel Corporation and is based on a Si/Si$_{0.7}$Ge$_{0.3}$ heterostructure \cite{Neyens2024}, where the quantum well is composed of isotopically-purified $^{28}$Si (800 ppm $^{29}$Si). Gate-defined QDs are formed by molding the local electrostatic potential with voltages applied to the surrounding gates. A reservoir of electrons is hosted under the accumulation gate AC0 from which the QDs are populated. Changes in charge occupation are sensed by a single electron transistor (SET) located at a dedicated QD on the top half of the device opposite the qubit QDs and isolated by a center screening gate. The dot pitch is $\sim$120 nm, the Si well thickness is $\sim$5 nm, and the SiGe barrier thickness is $\sim$50 nm.

\subsection{Experimental setup}
We use a closed-loop dry cryostat Bluefors LD400 dilution refrigerator with a base temperature of 7 mK. The electron temperature is approximately 250 mK. We use a heterojunction bipolar transistor (HBT) operating at -1.056 V emitter bias on the readout SET line to perform cryogenic preamplification \cite{Curry2019}. DC voltage to the gates is supplied by a QDevil QDAC II and fast AC pulses are generated by a Keysight M3201 AWG with a 500 MHz sampling rate. DC and fast voltage pulses are combined through a cryogenic RC bias tee at the PCB. Each individual spin state measurement is acquired over an average of 100 shots. We use one axis of a superconducting vector magnet aligned to the [110] crystal direction in the plane of the device, perpendicular to the direction of shuttling.

\section{References}
\bibliographystyle{naturemag}
\bibliography{refs}

\section{Acknowledgments}
We gratefully acknowledge Intel Corp. for providing devices and Matthew Curry, Lester Lampert and Nathan Bishop for helpful discussions on device operation and performance. We thank Charlotte I. Evans for offering advice on measurement acquisition software usage. We also thank Wayne M. Witzel and N. Toby Jacobson for fruitful conversations regarding the observed decay dynamics. Sandia National Laboratories is a multi-mission laboratory managed and operated by National Technology \& Engineering Solutions of Sandia, LLC (NTESS), a wholly owned subsidiary of Honeywell International Inc., for the U.S. Department of Energy’s National Nuclear Security Administration (DOE/NNSA) under contract DE-NA0003525. This written work is authored by an employee of NTESS. The employee, not NTESS, owns the right, title and interest in and to the written work and is responsible for its contents. Any subjective views or opinions that might be expressed in the written work do not necessarily represent the views of the U.S. Government. The publisher acknowledges that the U.S. Government retains a non-exclusive, paid-up, irrevocable, world-wide license to publish or reproduce the published form of this written work or allow others to do so, for U.S. Government purposes. The DOE will provide public access to results of federally sponsored research in accordance with the DOE Public Access Plan.

\section{Author contributions}
N.D.F. and R. M. J. designed the experiments. N. D. F. performed the experiments and analyzed the data. N. D. F., J. D. H., and R. M. J. discussed the results. J. D. H. and M. R. contributed to the measurement setup and performed initial shuttling demonstrations. N. D. F. wrote the manuscript with input from all co-authors. R. M. J. and D. R. L. supervised the project.


\section{Supplementary Information}
\setlength{\tabcolsep}{12pt}

\begin{table*}[t!]
\begin{tabular}{rllllll}
Step & Description                               & $V_\mathrm{P0}$ & $V_\mathrm{B1}$ & $V_\mathrm{P1}$ & $V_\mathrm{B2}$ & $V_\mathrm{P2}$ \\ \hline 
1           & reset (3,0,0)                             & 1.94            & 0.819           & 1.487           & 0.81            & 1.48            \\
2           & load (4,0,0)                              & 1.9696          & 0.819           & 1.487           & 0.81            & 1.48            \\
3           & roll                                      & 1.9625          & 0.819           & 1.5335          & 0.81            & 1.48            \\
4           & pulsed barrier B1 roll                       & 1.9305          & 0.869           & 1.5097          & 0.81            & 1.48            \\
5           & transfer (4,0,0) - (3,1,0)    & 1.879           & 0.869           & 1.5477          & 0.81            & 1.49            \\
6           & pulsed barrier shuttle prep               & 1.879           & 0.869           & 1.433           & 1.01            & 1.426           \\
7           & shuttle position (3,1,0)                  & 1.879           & 0.869           & 1.473           & 1.01            & 1.386           \\
8           & shuttle position (3,0,1)                  & 1.879           & 0.869           & 1.373           & 1.01            & 1.486           \\
9           & shuttle back (3,1,0)                      & 1.879           & 0.869           & 1.473           & 1.01            & 1.386           \\
10          & wait time                                 & 1.879           & 0.869           & 1.473           & 1.01            & 1.386           \\
11          & pulsed barrier B2 return roll                     & 1.879           & 0.869           & 1.433           & 1.01            & 1.426           \\
12          & close shuttling barrier & 1.879           & 0.869           & 1.5477          & 0.81            & 1.49            \\
13          & PSB window return roll                        & 1.9305          & 0.869           & 1.5097          & 0.81            & 1.48            \\
14          & measure in latched PSB                    & 1.9705          & 0.819           & 1.5407          & 0.81            & 1.48           
\end{tabular}
\caption{Pulse levels for a simple $N=1$ shuttling experiment. Repeated shuttles are implemented by repeating steps 7-9.}
\label{tab:shuttling_pulse_levels}
\end{table*}

\setlength{\tabcolsep}{6pt}
\begin{table*}[h]
\begin{tabular}{l|l|l|l|l|l|l}
\multicolumn{1}{c|}{fit parameter} & \multicolumn{1}{c|}{(3,1,0) at 0 mT} & \multicolumn{1}{c|}{(3,0,1) at 0 mT} & \multicolumn{1}{c|}{transition at 0 mT} & \multicolumn{1}{c|}{(3,1,0) at 50 mT} & \multicolumn{1}{c|}{(3,0,1) at 50 mT} & \multicolumn{1}{c}{transition at 50 mT} \\ \hline
$A$ ($\ket{S}$ \%)                  & $0.325 \pm 0.002$                    & $0.354 \pm 0.002$                    & $0.318 \pm 0.002$                       & $0.342 \pm 0.009$                     & $0.330 \pm 0.003$                     & $0.321 \pm 0.003$                        \\
$T_2^*$ ($\mu$s)                    & $4.60 \pm 0.03$                      & $3.86 \pm 0.01$                      & $5.12 \pm 0.03$                         & $3.82 \pm 0.12$                       & $4.09 \pm 0.03$                       & $1.08 \pm 0.01$                          \\
$\alpha$ (a. u.)                         & $1.92 \pm 0.03$                          & $2.16 \pm 0.03$                          & $2.10 \pm 0.03$                             & $0.90 \pm 0.06$                       & $2.12 \pm 0.06$                       & $1.14 \pm 0.02$                          \\
$f$ (MHz)                           & -                                    & -                                    & -                                       & $2.4031 \pm 0.0008$                   & $2.0356 \pm 0.0007$                   & $2.3516 \pm 0.0010$                      \\
$y_0$ ($\ket{S}$ \%)                & $0.374 \pm 0.001$                    & $0.408 \pm 0.001$                    & $0.399 \pm 0.001$                       & $0.473 \pm 0.001$                     & $0.481 \pm 0.001$                     & $0.4640 \pm 0.0001$                     
\end{tabular}
\caption{Fit parameters and uncertainties for Ramsey experiment shown in Fig.~\ref{fig:Fig2}c.}
\label{tab:Ramsey_fit_params}
\end{table*}

\subsection{Barrier pulsing}
For barrier-assisted bucket brigade shuttling we pulse the B2 barrier gate to increase the coupling between QD1 and QD2 and ensure adiabatic spin transfer. Supplementary Fig.~\ref{fig:FigS_dB2} shows a series of singlet return probability raster scans in the P1-P2 voltage space (left column), accompanied by linear detuning ramps of $V_\mathrm{P1}$ and $V_\mathrm{P2}$ (ramped in opposite directions) from (3,1,0) to (3,0,1) over time (right column). Nominally $V_\mathrm{B2}^0 = 0.81$ V before the transfer, and dB2$~=V_\mathrm{B2}-V_\mathrm{B2}^0$ V. For high barrier (small dB2, first row), the electron spin is transferred to (3,0,1) with low fidelity evidenced by the decrease in oscillation amplitude. The appearance of multiple $S$-$T_0$ oscillation frequencies in the first three rows with dB2 = 0.100, 0.125, 0.150 V (panels f.-h.) suggests the possibility of leakage into excited valley-orbit states \cite{Yang2012} during transfer. For dB2 = 0.175, 0.200 V in the last two rows, the $\ket{S}$ return probability is similar for both charge configurations, indicating the electron spin is transferred with minimal coherence loss while preserving the individual character of the QDs. The quickly dephasing singlet at the charge transition in panels f.-j. suggests the electron is strongly hybridized in both QDs. The criteria used to select dB2 for the experiments were singlet visibility preservation during the transfer and clear distinction of the charge transition, which are optimal in the case of dB2 = 0.200 V.

\subsection{Characterizing $\Delta g$ and the interdot charge transition}
We expect the rotation frequency $f$ to increase linearly with $B_0$ at a rate $\Delta g_{i,j}\mu_\mathrm{B}/h$ defined for each pair of QD$i$ and QD$j$, where $h$ is the Planck constant from the expression $E=hf$. We measure the singlet return probability as a function of wait time in each charge configuration (3,1,0) and (3,0,1), while an in-plane $B_0$ oriented along the [110] crystallographic axis is ramped from 0 to 300 mT. Supplementary Fig.~\ref{fig:FigS_Bramp-FFT} shows the raw data and normalized fast Fourier transforms, demonstrating $S$-$T_0$ rotation frequency increasing linearly with $B_0$ at lower fields for both QD0-QD1 and QD0-QD2 pairs, as expected. Linear fits to the data result in distinct slopes of 42.47 $\pm$ 0.21 MHz/T for QD0-QD1, and 36.26 $\pm$ 0.71 MHz/T for QD0-QD2 from which we estimate $\Delta g$-factors of $|\Delta g_{0,1}|=0.0030$ and $|\Delta g_{0,2}|=0.0026$. The observation of rapid $S$-$T_0$ rotations and divergence in frequency for both charge configurations near $B_0 =$ 190 mT suggests the presence of strong spin-valley coupling in QD0, the dot common to both configurations \cite{Jock2022}. Converting $B_0$ to energy using $E=g\mu_\mathrm{B}B_0$ and using $g\approx2$ for Si, this value corresponds to a valley splitting in QD0 of $\sim$21 $\mu$eV.
The appearance of a second divergence near $B_0=0.3$ T could be the emergence of a weakly coupled higher-energy valley state that is more pronounced in QD1 than in QD2, corresponding to a valley splitting energy $\gtrsim35$ $\mu$eV in QD1 and presumably even higher in QD2. Nonetheless, for the low fields used in this work ($B_0=0$ to 50 mT), we expect the influence of these unexplored features to be minimal.

We analyze the nature of the interdot charge transition in closer detail in Supplementary Fig.~\ref{fig:FigS_detuning-waittime} by transferring the prepared qubit state between QD1 and QD2 and observing the singlet return probability over time. Supplementary Fig.~\ref{fig:FigS_detuning-waittime}a. shows measured $S$-$T_0$ rotations in both charge sectors, and Supplementary Fig.~\ref{fig:FigS_detuning-waittime}b. shows the results of dephasing sine (Eq.~\ref{eq:dephase_sine}) fits to each time-trace in the data. We observe a clear distinction in $T_2^*$ based on detuning (${\sim}6$ $\mu$s in (3,1,0), ${\sim}4.5$ $\mu$s in (3,0,1) and ${\sim}1$ $\mu$s at the charge transition), as well as a gradual decrease in $S$-$T_0$ rotation frequency (from ${\sim}2.41$ to ${\sim}2.12$ MHz) as detuning is ramped from (3,1,0) to (3,0,1) that appears to flatten off farther from the charge transition. We note that the $T_2^*$ values obtained using this fit are slightly longer than the values reported for the data in Fig.~\ref{fig:Fig2}c, and conclude that this is expected of data averaged well below the ergodic limit \cite{Madzik2020} (see discussion on Supplementary Fig.~\ref{fig:FigS_T2star} for more information). Overall, these observations demonstrate that the electron is well-isolated in QD1 or QD2 because the qubit is shown to experience unique $S$-$T_0$ rotation frequencies due to unique $\Delta g$ pertaining to each QD.

\subsection{Tunnel coupling and nonadiabatic transition probability}

The tunnel coupling ($t_c$) between QD1 and QD2 during shuttling was not measured explicitly, but we can estimate $t_c$ based on a proxy measurement similar to \cite{Wang2013}. Supplementary Fig.~\ref{fig:FigS_detuning-waittime} shows Ramsey-like measurements of $S$-$T_0$ rotations as the detuning between QD1 and QD2 is ramped through the charge transition. We infer the relative probability of occupying QD1 or QD2 from the rotation frequency since it provides a measure of $\Delta g$, and thus the local QD spin-orbit environment as the electron wavefunction transitions from QD1 to hybridized at the transition to QD2. We fit the data to the following equation 
\begin{equation}
    P_\mathrm{(QD1,QD2)}=\frac{1}{2}\left(1-\frac{\epsilon}{\Omega}\mathrm{tanh}\frac{\Omega}{2k_BT_e}\right)
\label{eq:tc}
\end{equation}
where $\Omega=\sqrt{\epsilon^2+4t_c^2}$. We use electron temperature $T_e \approx 250$ mK measured for our device and fit the equation to the range of extracted rotation frequency normalized to the range [0,1] (see Supplementary Fig.~\ref{fig:FigS_detuning-waittime}b). We calculate the detuning $\epsilon$ between QD chemical potentials $\mu$ by applying the QD lever arms $\alpha$ to $V_\mathrm{P1}$ and $V_\mathrm{P2}$ in the following manner:
\begin{equation}
\begin{gathered}
    \epsilon = \mu_2 - \mu_1 \\
    \mu_1 = \alpha_{11}V_\mathrm{P1} + \alpha_{21}V_\mathrm{P2} \\
    \mu_2 = \alpha_{12}V_\mathrm{P1} + \alpha_{22}V_\mathrm{P2} \\
\end{gathered}
\end{equation}
where $\alpha_{11}$ ($\alpha_{22}$) $\approx70$ $\mu$eV/mV is the lever arm of $V_\mathrm{P1}$ ($V_\mathrm{P2}$) on QD1 (QD2), and $\alpha_{12}\approx17~\mu$eV/mV ($\alpha_{21}\approx18~\mu$eV/mV) is the lever arm accounting for cross-capacitance of $V_\mathrm{P1}$ ($V_\mathrm{P2}$) on QD2 (QD1) extracted from the slopes of the charge transitions.
The resulting tunnel coupling between QD1 and QD2 during shuttling is $t_c$ = 52 GHz or around 200 $\mu$eV.

Further, we calculate the probability of nonadiabatic Landau-Zener transitions during the shuttle between QD1 and QD2 \cite{Nichol2015}:
\begin{equation}
    P_\mathrm{LZ}=1-\mathrm{exp}\left( -\frac{2\pi(2t_c)^2}{\hbar dE/dt}\right)
\label{eq:LZ_probability}
\end{equation}
\\
$dE/dt = 183$ keV/s is the energy velocity across the charge transition considering the ramp time of 10 ns and voltage $\sim$25 mV. This results in a probability of non-adiabatic LZ transfer of effectively zero.

%
\renewcommand{\figurename}{Supplementary Figure}
\setcounter{figure}{0}

\begin{figure*}
\centering
\includegraphics[width=0.75\textwidth]{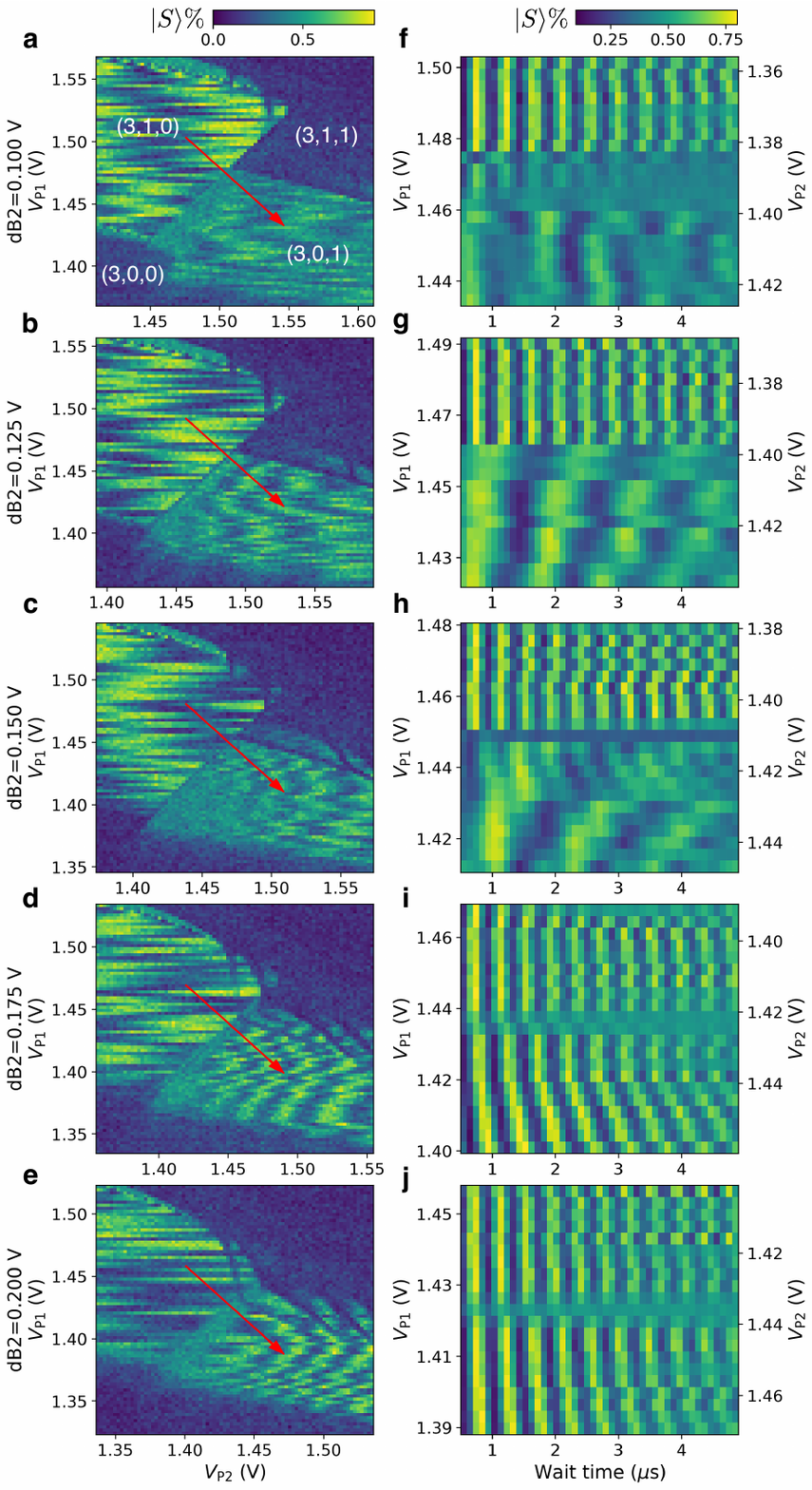}
	\caption{Barrier-enhanced coherent spin transfer. \textbf{a.-e.} Singlet return probability raster scans of the charge space between (3,1,0) and (3,0,1) states for a range of barrier pulse voltages dB2 $= 0.1-0.2$ V used during the shuttle. The wait time in each panel is 6 $\mu$s. \textbf{f.-j.} Singlet return over time as the detuning is ramped from (3,1,0) to (3,0,1) along the trajectory defined by red arrow in the corresponding raster plot shown in the panel directly to the left. $B_0=50$ mT and the detuning ramp time is 30 ns in all measurements shown.}
\label{fig:FigS_dB2}
\end{figure*}

\begin{figure*}
\centering
\includegraphics[width=0.85\textwidth]{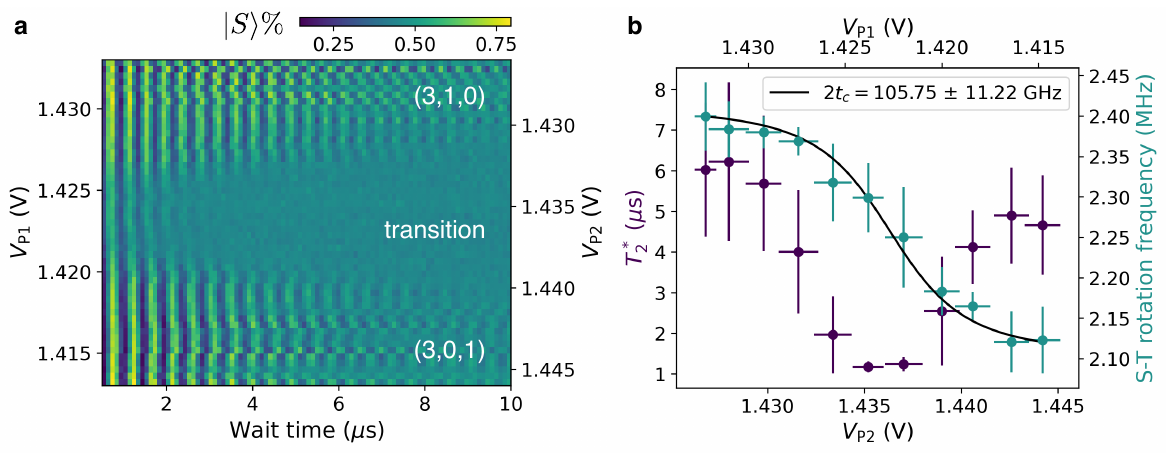}
	\caption{Shuttling from QD1 to QD2. \textbf{a.} Singlet return probability over time as the detuning is ramped from (3,1,0) to (3,0,1) with $B_0=50$ mT. \textbf{b.} Averaged fit parameters of $T_2^*$ dephasing time (left y-axis) and $S$-$T_0$ rotation frequency (right y-axis) from a dephasing sine model (Eq.~\ref{eq:dephase_sine} with $\alpha=2$) fit to the data in \textbf{a.} The black curve represents a tunnel coupling fit to the rotation frequency parameters using Eq.~\ref{eq:tc}}.
\label{fig:FigS_detuning-waittime}
\end{figure*}

\begin{figure*}
\centering
\includegraphics[width=0.75\textwidth]{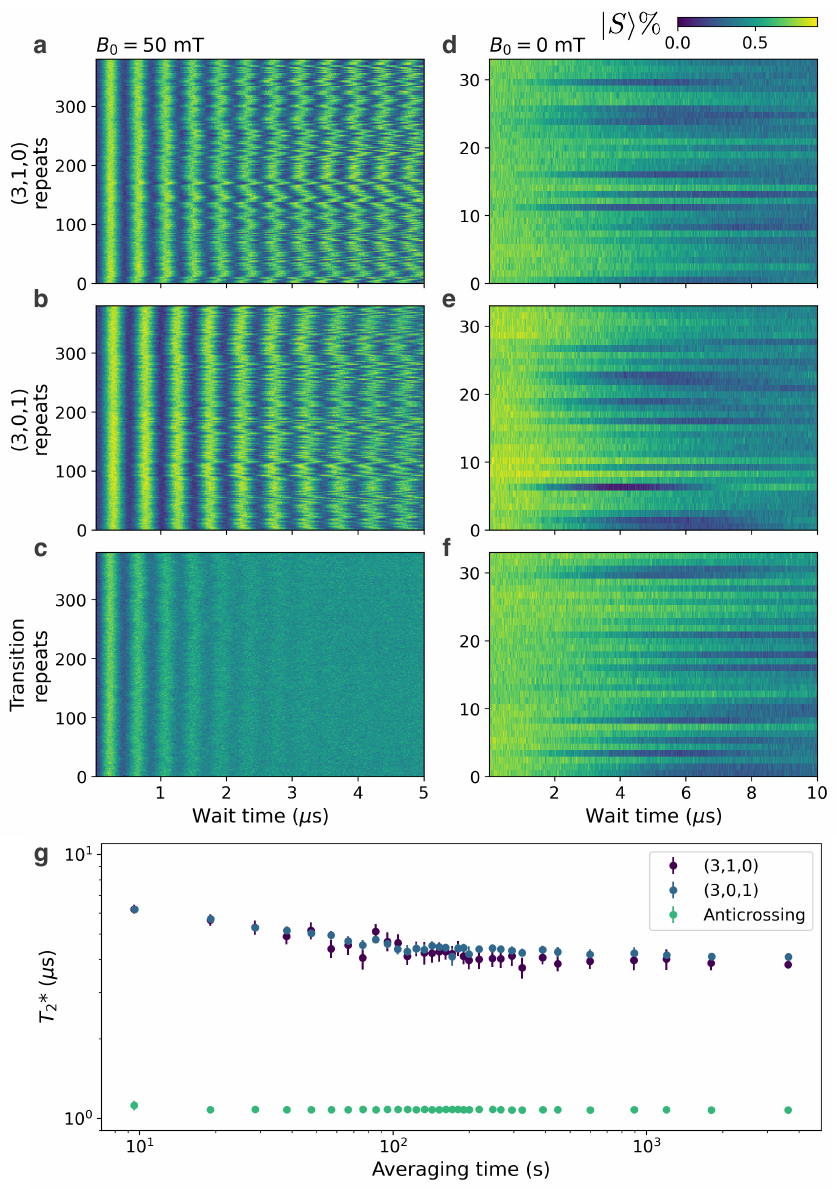}
	\caption{$T_2^*$ repeated measurements spanning 1 hour. \textbf{a.-c.} Repeated time traces of singlet return probability for (3,1,0), (3,0,1) and the charge transition, respectively, at $B_0=50$ mT. \textbf{d.-f.} Repeated time traces of singlet return probability for (3,1,0), (3,0,1) and the charge transition, respectively, at $B_0\approx0$ mT. \textbf{g.} Gaussian $T_2^*$ fits to each line trace over time in \textbf{a.-c.} averaged over an increasing number of line traces. The averaging time on the horizontal axis is derived from the lab time accumulated over the course of data acquisition, including dead time to read out and send information to the computer. $T_2^*$ fits approach the ergodic limit after about 10 minutes of averaging.}
\label{fig:FigS_T2star}
\end{figure*}

\begin{figure*}
\centering
\includegraphics[width=0.8\textwidth]{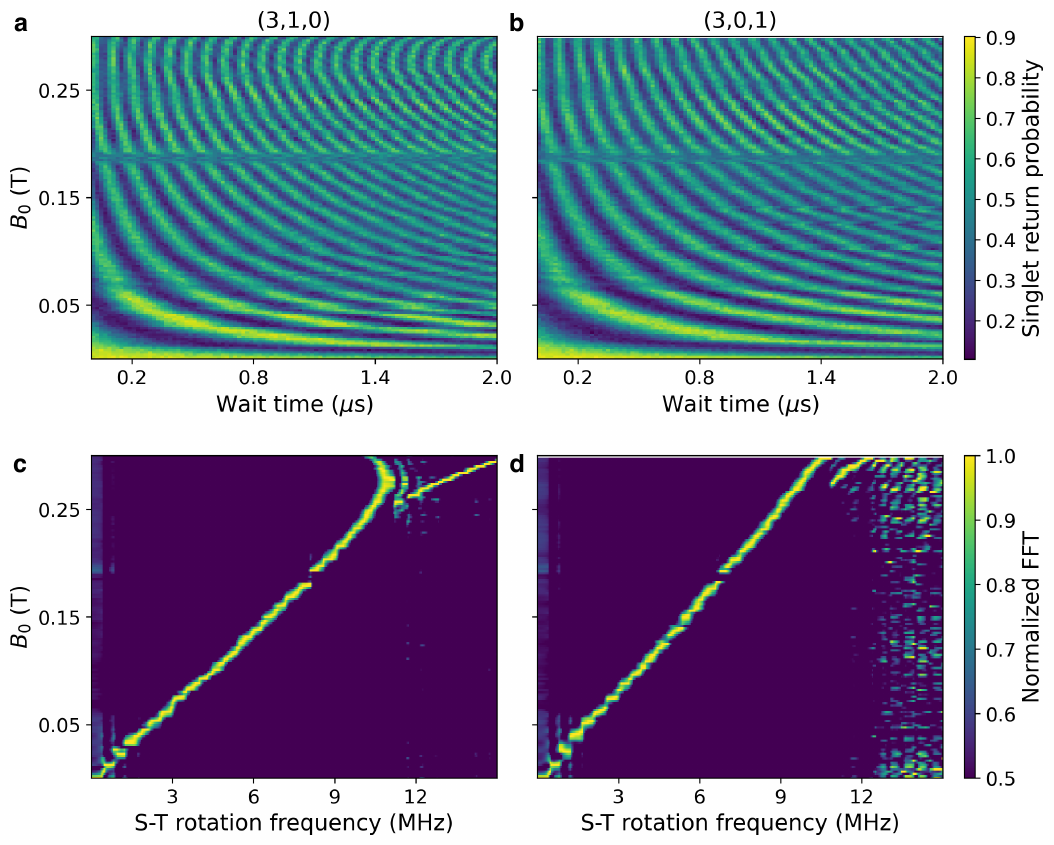}
	\caption{$S$-$T_0$ oscillations in an external magnetic field for (3,1,0) and (3,0,1) charge configurations. \textbf{a.-b.} Singlet return probability as $B_0$ is ramped. The qubit is held for a variable wait time in each of (3,1,0) and (3,0,1). \textbf{c.-d.} Corresponding normalized FFTs to the data shown in \textbf{a.,b.} showing $S$-$T_0$ qubit rotation frequency.}
\label{fig:FigS_Bramp-FFT}
\end{figure*}

\end{document}